\documentclass[conference]{IEEEtran}
\IEEEoverridecommandlockouts
\usepackage{cite}
\usepackage{amsmath,amssymb,amsfonts}
\usepackage{algorithm}
\usepackage{algpseudocode}
\usepackage{graphicx}
\usepackage{textcomp}
\usepackage{xcolor}
\usepackage{enumitem}
\usepackage{comment}
\def\BibTeX{{\rm B\kern-.05em{\sc i\kern-.025em b}\kern-.08em
    T\kern-.1667em\lower.7ex\hbox{E}\kern-.125emX}}
\begin{document}

\title{Quantum Markov Chain Monte Carlo for Cosmological Functions
}

\author{\IEEEauthorblockN{Giuseppe Sarracino}
\IEEEauthorblockA{\textit{OACN,}
\textit{INAF,}
Via Moiariello 16, 80131, Napoli, Italy \\
giuseppe.sarracino@inaf.it} 
\and
\IEEEauthorblockN{Vincenzo Fabrizio Cardone}
\IEEEauthorblockA{\textit{OAR,}
\textit{INAF,}
Via Frascati 33, 00078, Monteporzio Catone, Italy \\
vincenzo.cardone@inaf.it} 
\and
\IEEEauthorblockN{Roberto Scaramella}
\IEEEauthorblockA{\textit{OAR,}
\textit{INAF,}
Via Frascati 33, 00078, Monteporzio Catone, Italy \\
roberto.scaramella@inaf.it} 
\and
\IEEEauthorblockN{Giuseppe Riccio}
\IEEEauthorblockA{\textit{OACN,}
\textit{INAF,}
Via Moiariello 16, 80131, Napoli, Italy \\
giuseppe.riccio@inaf.it} 
\and
\IEEEauthorblockN{Andrea Bulgarelli}
\IEEEauthorblockA{\textit{OAS,}
\textit{INAF,}
Via Pietro Gobetti 93/3, 40129, Bologna, Italy \\
andrea.bulgarelli@inaf.it} 
\and
\IEEEauthorblockN{Carlo Burigana}
\IEEEauthorblockA{\textit{IRA,}
\textit{INAF,}
Via Pietro Gobetti 101, 40129, Bologna, Italy  \\
carlo.burigana@inaf.it} 

\and
\IEEEauthorblockN{Luca Cappelli}
\IEEEauthorblockA{\textit{OATs,} 
\textit{INAF,}
Via Tiepolo 11, I-34131, Trieste, Italy \\
luca.cappelli@inaf.it} 

\and

\IEEEauthorblockN{Stefano Cavuoti}
\IEEEauthorblockA{\textit{OACN,}
\textit{INAF,}
Via Moiariello 16, 80131, Napoli, Italy \\
stefano.cavuoti@inaf.it} 
\and
\IEEEauthorblockN{Farida Farsian}
\IEEEauthorblockA{\textit{OACT,}
\textit{INAF,}
Via S. Sofia 78, 95123, Catania, Italy \\
farida.farsian@inaf.it}
\and
\IEEEauthorblockN{Irene Graziotti}
\IEEEauthorblockA{\textit{OACN,}
\textit{INAF,}
Via Moiariello 16, 80131, Napoli, Italy \\
irene.graziotti@inaf.it}
\and
\IEEEauthorblockN{Massimo Meneghetti}
\IEEEauthorblockA{\textit{OAS,}
\textit{INAF,}
Via Pietro Gobetti 93/3, 40129, Bologna, Italy \\
massimo.meneghetti@inaf.it} 
\and
\IEEEauthorblockN{Giuseppe Murante}
\IEEEauthorblockA{\textit{OATs,} 
\textit{INAF,}
Via Tiepolo 11, I-34131, Trieste, Italy \\
giuseppe.murante@inaf.it} 
\and

\IEEEauthorblockN{Nicolò Parmiggiani}
\IEEEauthorblockA{\textit{OAS,}
\textit{INAF,}
Via Pietro Gobetti 93/3, 40129, Bologna, Italy \\
nicolo.parmiggiani@inaf.it} 

\and
\IEEEauthorblockN{Alessandro Rizzo}
\IEEEauthorblockA{\textit{OACT,}
\textit{INAF,}
Via S. Sofia 78, 95123, Catania, Italy \\
alessandro.rizzo@inaf.it} 
\and
\IEEEauthorblockN{Francesco Schillirò}
\IEEEauthorblockA{\textit{OACT,}
\textit{INAF,}
Via S. Sofia 78, 95123, Catania, Italy \\
francesco.schilliro@inaf.it} 
\and

\IEEEauthorblockN{Vincenzo Testa}
\IEEEauthorblockA{\textit{OAR,}
\textit{INAF,}
Via Frascati 33, 00078, Monteporzio Catone, Italy \\
vincenzo.testa@inaf.it} 

\and
\IEEEauthorblockN{Tiziana Trombetti}
\IEEEauthorblockA{\textit{IRA,}
\textit{INAF,}
Via Pietro Gobetti 101, 40129, Bologna, Italy  \\
tiziana.trombetti@inaf.it} 

}

\maketitle

\begin{abstract}
We present an implementation of Quantum Computing for a Markov Chain Monte Carlo method with an application to cosmological functions, to derive posterior distributions from cosmological probes. The algorithm proposes new steps in the parameter space via a quantum circuit whose resulting statevector provides the components of the shift vector. The proposed point is accepted or rejected via the classical Metropolis-Hastings acceptance method. The advantage of this hybrid quantum approach is that the step size and direction change in a way independent of the evolution of the chain, thus ideally avoiding the presence of local minima. The results are consistent with analyses performed with classical methods, both for a test function and real cosmological data. The final goal is to generalize this algorithm to test its application to complex cosmological computations.

\end{abstract}


\section{Introduction}
Quantum Computing (QC) is an emerging field that has been advancing rapidly in recent years, drawing substantial interest from many fields, among which the scientific community \cite{Steane_1998, AHARONOV_1999, ladd_2010, Li_2020}. It is based on the idea of using fundamental concepts of Quantum Mechanics, such as Entanglement and Superposition, in conjunction with Computer Science to define a novel approach to both hardware and software implementations. For the latter, the paradigm is modified with respect to classical algorithms in the sense that the fundamental unit becomes the Qubit, which can exist in a superposition of states until a measurement is performed. 

Theoretically, QC has been studied since the end of the last century \cite{Feynman_1982,  Unruh_1995, DiVincenzo_1995, Narayanan_1996}. In this period, the first proper quantum algorithms were designed, finding cases in which a quantum advantage with respect to the classical counterparts has been demonstrated, at least in theory \cite{Grover_1998, Shor_1999}. In recent years, a remarkable improvement has been achieved from the hardware point of view, which has brought us to the first proper Quantum Computers \cite{Massarotti_2023}. This is the main reason why a significant effort is currently underway to find the so-called "Quantum Utility," i.e. practical applications for which QC performs better than classical algorithms. Such advantage can be defined as faster convergence, using fewer resources, or finding better results. For a recent review of quantum algorithms presented in the literature, see \cite{Montanaro_2016, Dalzell_2023}.

Among the different scientific fields in which interesting applications of QC could be found, astrophysics and cosmology are the ones we focus on. We live in an epoch of astronomical data richness, for which vast, high-quality data catalogs are at the disposal of the astronomical community, and strategies for efficiently searching and analyzing these datasets are becoming mandatory \cite{Taffoni_2020}. Examples of missions and instruments that have given us such remarkable datasets are Gaia \cite{Ripepi_2023}, the Sloan Digital Sky Survey (SDSS, \cite{York_2000}), and the Very Large Telescope (VLT, \cite{Lilly_2007}). These will be accompanied by data provided by novel instruments like Euclid \cite{Laureijs_2011, Scaramella_2022, Mellier_2024}, and the Vera C. Rubin Observatory \cite{Abell_2009}. Efficient and fast analyses have been performed with novel strategies like machine learning models \cite{Brescia_2015, Angora_2020, Sen_2022}  as well as the redesign of algorithms to employ high-performance computing (HPC) hardware as efficiently as possible. The idea is to understand if QC can be used in this context, looking for possible applications where Quantum Utility could be found.

In this paper, we focus our attention on the Bayesian inference of posterior probabilities for cosmological functions, regarding cosmological $\chi^2$ functions defined using Supernovae Type Ia (SNe Ia) and the Cosmic Microwave Background (CMB) radiation. Bayesian inference is widely used in a cosmological context to derive estimates of fundamental cosmological parameters, but given the size and complexity of the parameter space offered by the novel missions, it is becoming more and more time-consuming and computationally expensive. With this in mind, we build a Quantum Markov Chain Monte Carlo (QMCMC) algorithm, which proposes the new steps of the chain via quantum operations while evaluating the acceptance rate classically. In section 2, we briefly present the related literature on optimization and sampling problems solved via quantum algorithms, as well as the cosmological probes we have considered in our study. In section 3 we describe our algorithm, while we show our first results in section 4. Finally, conclusions are drawn in section 5.

\section{Overview of the Problem}

We present here the cosmological probes used in our analysis. SNe Ia are widely used as late-type probes of our universe because of their role as standard candles in the cosmological ladder \cite{Riess_1998, Abdalla_2022}. Indeed, it is possible to infer their intrinsic luminosity from a particular relation \cite{Phillips_1993} between observational features that are independent of the distance. From this, one can derive the luminosity distance defined as 

\begin{equation} \label{luminosity distance}
    d_{\rm L}(z)=(1+z)d_{\rm M}(z)~,
\end{equation}
where $d_{\rm M}(z)$ is the transverse comoving distance
\begin{equation} \label{comoving flat}
    d_{\rm M}(z)=\frac{c}{H_0} \int_0^z \frac{dz'}{E(z')}~,
\end{equation}
where $H_0$ is the Hubble Constant (defined in the rest of the analysis in Km/s/Mpc) and we model $E(z)$ as
\begin{equation} \label{E_z}
\begin{split}
    E(z) = \frac{H(z)}{H_0} = \\
    \sqrt{\Omega_M(1+z)^3 +\Omega_{\Lambda}(1+z)^{3(1+w)}}~.
\end{split}
\end{equation}
where we neglect the radiation term and assume the flatness of the Universe. $\Omega_{M}$ is the density associated with the matter component of the Universe, $\Omega_{\Lambda}$ is the "density" associated with the Dark Energy one, and $w$ is the equation of state parameter for a $w$CDM model, which becomes the standard $\Lambda$CDM cosmological model if $w=-1$. From this, one defines the distance modulus

\begin{equation}
    \mu_{th, SNe Ia}=m-M=5 \log (d_L)+25,
\label{muth}
\end{equation}
where $m$ is the apparent magnitude of the SN Ia, $M$ is its absolute magnitude, and the luminosity distance is expressed in Mpc. This is compared with the observed distance modulus, $\mu_{obs}$ of the SN Ia by defining a $\chi^2$ function
\begin{equation} \label{eq_chi2_SNe}
    \chi^2_{SNe Ia}= (\mu_{th}-\mu_{obs})^T \mathcal{C}_{SNe Ia}^{-1} (\mu_{th}-\mu_{obs})~,   
\end{equation}
 where $\mathcal{C}_{SNe Ia}^{-1}$ is the inverse of the covariance matrix. In our case, we have used the Pantheon+ set of SNe Ia, which is a compilation of 1701 light curves gathered from 1550 different SNe Ia \cite{Scolnic_2022, Brout_2022}.

 For the CMB, instead, we recall that we are dealing with an early-type probe, allowing us to infer cosmological parameters from the observation of the early phases of the Universe in a cosmology-dependent way. For our analysis, we have used the Temperature-Temperature (TT) Power Spectrum provided by the latest CMB measurements by the Planck mission \cite{Planck2020}. This spectrum has been compared with the theoretical one derived from tools like \texttt{CAMB} \cite{Lewis_2000}, and \texttt{PICO} \cite{Fendt_2007, Fendt_2007b}, by defining again a $\chi^2$ function which quantifies the difference between the model and the observations in the usual way. The CMB constrains the $\Lambda$CDM model exceptionally well, allowing us to perform Bayesian computation on 5 different parameters contemporaneously, considering only the TT spectrum, which is a limited section of the entire dataset provided by Planck. Indeed, we have computed for the CMB in our analysis $\Omega_M$, $H_0$, $\omega_B$, $n_s$, and $A_s$, where  $\omega_B= \Omega_b h^2$ is the Baryon density scaled by the normalized Hubble parameter, $n_s$ is the scalar spectral index, and $A_s$ is the amplitude of the scalar perturbations. For more details, see \cite{Planck2020}.

 We now introduce the MCMC method \cite{Hastings_1970}, which is widely employed to derive posterior probabilities of cosmological likelihoods. This is a sampling method used in Bayesian statistics to converge to a given posterior distribution after a "prior" has been provided. The acceptance rate computed at each step of the chain is usually the Metropolis-Hastings, defined as 
 \begin{equation} \label{Metropolis-Hasting}
     \alpha = \min\left(1, \frac{\pi(\theta') q(\theta \mid \theta')}{\pi(\theta) q(\theta' \mid \theta)}\right),
 \end{equation}
 where $\theta'$ defines the proposed set of the chain, $\theta$ the current one, $\pi(\theta)$ the target distribution, and $q(\theta' \mid \theta)$ the proposed distribution, which cancels out in the computations if it is symmetric. This parameter is confronted with a number randomly generated between 0 and 1. If $\alpha$ is bigger than this random number, the step is accepted, and the chain moves to the new position; otherwise, the step is rejected and the chain remains in the previous one. This is repeated until convergence is reached. This can be checked in various ways. Those considered in our work are the Gelman-Rubin $\hat{R}-1$ convergence criterion \cite{Gelman_1992}, and the autocorrelation time $\tau$ \cite{Foreman_Mackey_2013}. In literature, one finds many tools for Bayesian computations following the MCMC methodology, also focused on cosmological applications, like \texttt{COBAYA} \cite{Torrado_2021}, or \texttt{emcee} \cite{Foreman_Mackey_2013}. The main issue is that, as previously mentioned, because of the richness and complexity of the new observed data and the refinement of the cosmological models, one has to fit more and more free parameters for complex likelihoods regarding very vast data-sets, and thus the classical Bayesian method is becoming more and more resource-heavy and time-consuming, especially for cosmological probes such as weak lensing and galaxy clustering \cite{Scaramella_2022}, which require taking into account a huge amount of free parameters. This is the reason why we aim to investigate if it is possible to formulate and solve this problem with the help of QC.

 From the QC point of view, optimization problems have been solved with quantum algorithms, in particular for combinatorial optimizations \cite{farhi_2014}, which may also be translated into problems solved via Quantum Annealing \cite{Rajak_2022}, or for finding the ground states of given Hamiltonians \cite{Cerezo_2021}. Quantum Genetic Algorithms have also been developed with binary encoding to find the minima of continuous functions \cite{Acampora_2021}. Regarding a Quantum version of the MCMC, a remarkable implementation can be found in \cite{Layden_2023}, which translates the sampling process into an Ising configuration and the related Boltzmann distribution, for which it proposes each step via quantum computations and then evaluates the acceptance rate and thus the merit function classically. This has given promising results, even if some discussion on this outcome has arisen \cite{Orfi_2024}. Other recent studies explore different applications of quantum sampling problems, among which are quantum Monte Carlo \cite{Mansky_2023}, quantum-enhanced MCMC sampling in physical systems \cite{Nakano_2024}, and quantum annealing-enhanced MCMC for molecular simulations \cite{Arai_2025}. We note that the fields of application differ substantially from the high-dimensional cosmological inference targeted in this work.

\section{Our QMCMC algorithm}
The idea is to propose steps via a quantum circuit, and then evaluate the acceptance rate classically. We proceed as follows:

\begin{figure*}
    \centering
\includegraphics[width=0.9\hsize]{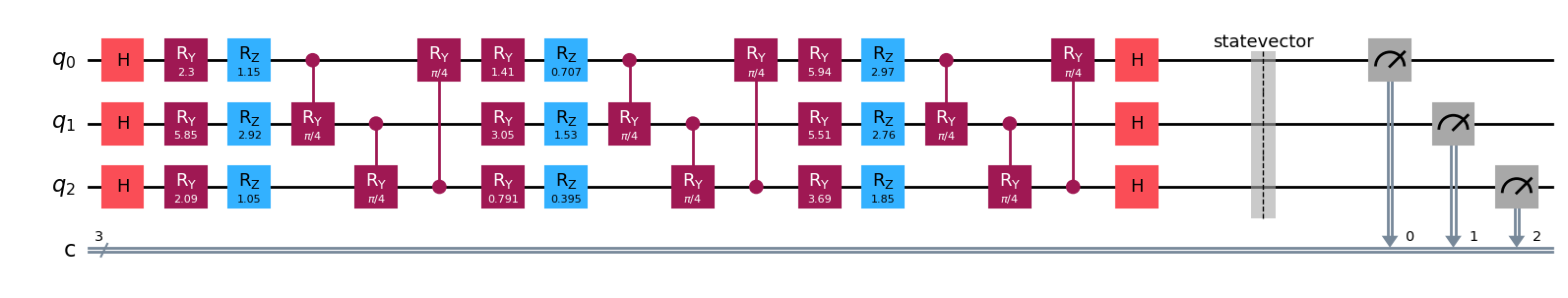}
\caption{The Quantum Circuit used in our algorithm to propose the new step in the chain.}
\label{Fig_Quantum_Shift}
\end{figure*}

    \textbf{1)} We first define the hyperparameters of the algorithm. i.e. the number of dimensions $d$, the bounds in which the chains can move, the initial step size $\mathbf{i}$, the starting point of the chains, the number of chains, the convergence criteria (and how many times they are evaluated during the run), the eventual prior contribution, and the number of burn-in steps. The algorithm can define uniform and Gaussian priors for the parameters.
    
    \textbf{2)} Then, we propose a new step for the chains using the quantum circuit shown in Fig. \ref{Fig_Quantum_Shift}. The circuit is composed of 3 layers of rotational gates on the $y$ and $z$ axes as well as conditional rotations on the $y$ axis linking all the qubits, and starts and ends with Hadamard gates. It has been built as such to not favour any particular quantum state, ensuring an even and unbiased exploration of the parameter space, while creating entanglement among the qubits. From this circuit, one computes the shifts $\mathbf{s}$ for the QMCMC as follows:
    \begin{equation} \label{step}
        \mathbf{s}=\mathbf{i} \cdot \operatorname{Re}(\mathbf{v})\cdot f(\operatorname{Im}(\mathbf{v}))
    \end{equation}
    where $\mathbf{v}$ is the statevector derived from the quantum circuit and $f$ is a step function depending on the components of the imaginary part of $\mathbf{v}$, multiplying or dividing the step size according to their values. Given that each component of the statevector influences the proposed step for a given dimension, the number of qubits necessary to build this quantum circuit scales with $\log_2(d)$,  while the depth scales as follows: $2+n_{l}\cdot (2+\log_2(d))$, where $n_{l}$ is the number of layers which constitute the circuit.
    
    \textbf{3)} Once the new step is proposed from the quantum circuit, the acceptance rate is computed classically using Eq. \ref{Metropolis-Hasting}, where $\pi$ is the objective function we are computing. Then the step is accepted or rejected following the rule shown in the previous section.
    
    \textbf{4)} The algorithm then iterates these two steps until convergence is reached. As convergence criteria, it currently uses both the autocorrelation time $\tau$ and the Gelman-Rubin statistics, but it can be easily adjusted to consider only one of these. More specifically, $\tau$ checks how many steps are necessary for the MCMC samples of a single chain to be non-correlated to the previous steps, while the $R-1$ statistics evaluates the variance within and between multiple MCMC chains to check if they have all converged to the same distribution. These convergence criteria are computed every $n$ steps of the chains, with $n$ chosen at the beginning of the run.
    
    \textbf{5)} Once the chains have converged, they are saved to derive the Bayesian contours, so that a comparison with classical algorithms is performed. 

\begin{algorithm} \label{Alg_QMCMC}
\caption{QMCMC Algorithm}
\begin{algorithmic}[1]
\State \textbf{Initialize:} 
\Statex \quad - Number of dimensions $d$
 \quad - Parameter bounds
 \quad - Initial step size $\mathbf{i}$
 \quad - Initial point(s) for chains
 \quad - Number of chains
 \quad - Convergence criteria ($\tau$, $R-1$) and frequency $n$
 \quad - Prior (facultative) 
 \quad - Number of burn-in steps

\While{convergence not reached}
    \For{each chain}
        \State Generate quantum statevector $\mathbf{v}$
        \State Compute shift as in Eq. \ref{step}.
        \State Propose new point using step $\mathbf{s}$
        \State Evaluate acceptance probability via Metropolis-Hastings, accept or reject accordingly 
    \EndFor
    \If{step $\mod$ $n$ == 0}
        \State Compute convergence diagnostics:
        \Statex \quad - Autocorrelation time $\tau$
        \Statex \quad - Gelman-Rubin $R-1$ statistic
    \EndIf
\EndWhile

\State Save chains for posterior analysis (e.g., Bayesian contours)

\end{algorithmic}
\end{algorithm}

The overall algorithm is summarized in \ref{Alg_QMCMC}. The main difference between this QMCMC and the classical MCMC methods found in the literature is in how the step is proposed. Indeed, given that both the real and imaginary parts of the statevector are used, the exploration of the parameter space offers a variability that can be obtained only via quantum operations. This variability is also independent of the previous steps in the chains, thereby avoiding possible internal correlations. From the scalability point of view, given that the number of qubits scales with  $\log_2(d)$, the size of the circuit stays fairly limited, in principle, even for complex cosmological problems. The depth also scales in the same way as the dimensions, plus the component given by the number of layers of the circuit, which can be regulated accordingly. Up to now, our algorithm has been tested with the Qiskit emulator.\footnote{https://docs.quantum.ibm.com}\cite{Javadi-Abhari_2024}, via an 12th Gen Intel(R) Core(TM) i5-1235U (1.30 GHz) with 10 cores and 16 GBs of RAM. We acknowledge that testing on real quantum hardware is necessary in future works to further assess the practical reliability of our QMCMC, taking into account potential limitations such as qubit connectivity, gate errors, and noise accumulation in deeper circuits.

\section{Results}

\begin{figure*}
    \centering
\includegraphics[width=0.5\hsize]{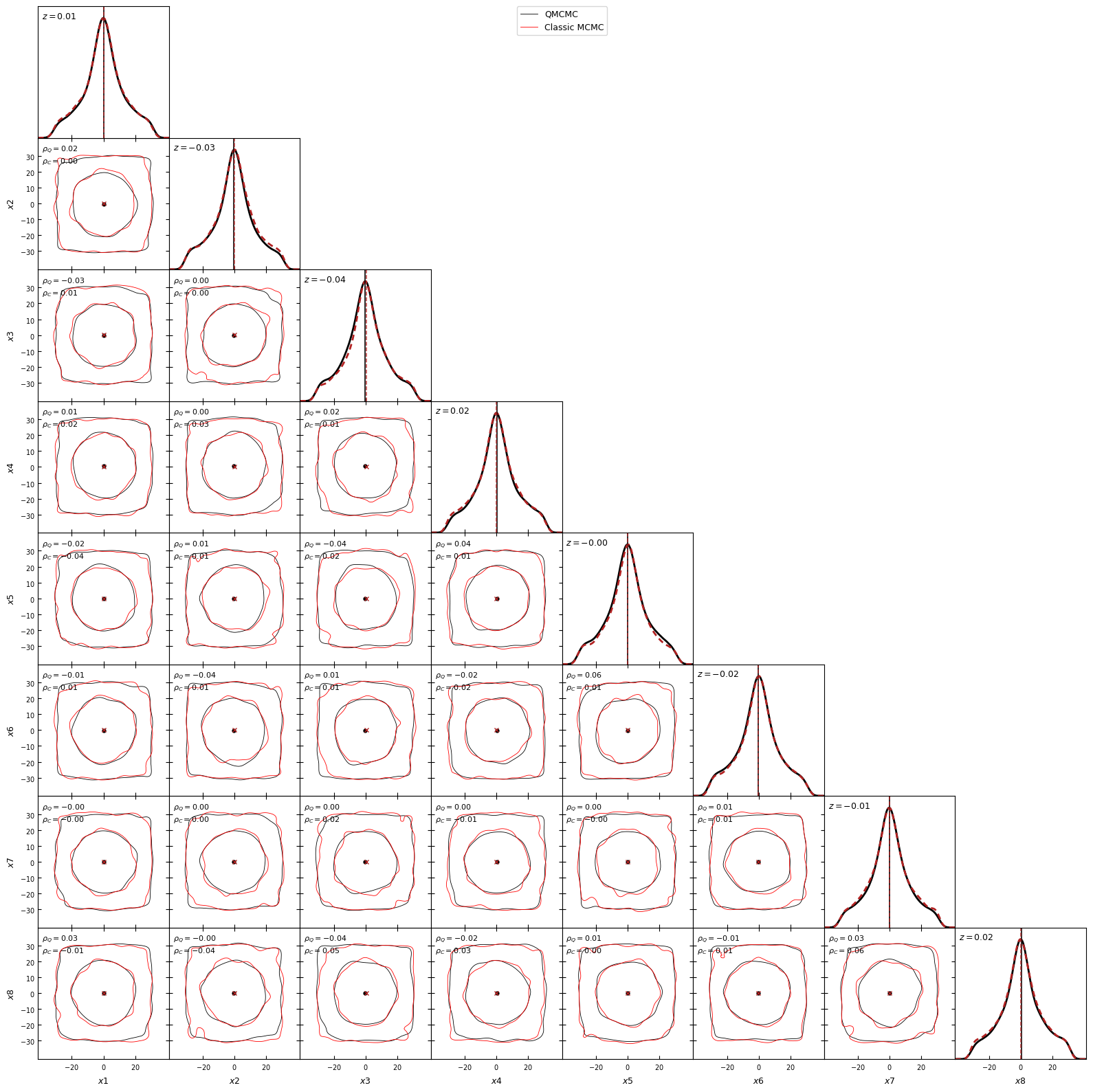}
\caption{$68\%$ and $95\%$ Contour plots for the Ackley test function with $d=8$, derived from our QMCMC algorithm and the classical MCMC \texttt{emcee} tool. Here, also the $z$-score and $\rho$ for each panel are shown.}
\label{Fig_Ackley_results}
\end{figure*}

We now present the first results we have obtained with our QMCMC. As the convergence check, for our QMCMC we ask the chains to be 50 times the mean $\tau$ for the parameters, and $R-1<0.05$. We consider $2000$ steps for the burn-in phase, while the convergence checks have been computed every $500$ steps after the burn-in. In all cases, we consider as the classical counterpart the results found by using the \texttt{emcee} tool, which uses a different sampling method and checks the convergence via $\tau$. In each comparison plot of this work, we show the Pearson correlation coefficient $\rho$, and the $z$-score $\frac{\mu_Q - \mu_C}{\sqrt{\sigma_Q^2 + \sigma_C^2}}$ to assess the difference between the classical and quantum results. We start from the Ackley test function \cite{Ackley_1987}
\begin{equation} \label{Ackley}
\begin{split}
    f(x_1, x_2, \dots, x_n) =  \\
    -a \exp\left(-b \sqrt{\frac{1}{n} \sum_{i=1}^{n} x_i^2}\right) - \exp\left(\frac{1}{n} \sum_{i=1}^{n} \cos(c x_i)\right) + a + e^1,
\end{split}
\end{equation}
where $a=20$, $b=2$, and $c=2\pi$. This function presents a global minimum in $f(0, 0, \dots, 0) = 0$ and several local minima around it, thus making it an ideal test bed for optimization and sampling algorithms. The results are shown in Fig. \ref{Fig_Ackley_results}. 

Here, we used $d=8$ and no prior has been given (corresponding to an infinite uniform prior, thus to an uninformative prior). For QMCMC, we used 50 chains. The comparison shows that the two algorithms find the same mean, the shapes of the contours almost overlap, the $z$-score is very close to 0, displaying a very good quantitative match between the results, and $\rho$ is almost 0 in all cases, showing consistency also in the correlation shape. 

For the QMCMC run, the chains have converged after 3500 steps (1500 if only the $R-1$ was considered). At the end of the run, the acceptance rate was $48\%$, while the mean effective sample size (ESS, computed as the mean of the number of steps per parameter divided by the $\tau$ for that parameter) is $2529$, indicating low autocorrelation and high sampling efficiency. The elapsed time for the run is $792$ seconds.

\begin{figure*}
    \centering
\includegraphics[width=0.45\hsize]{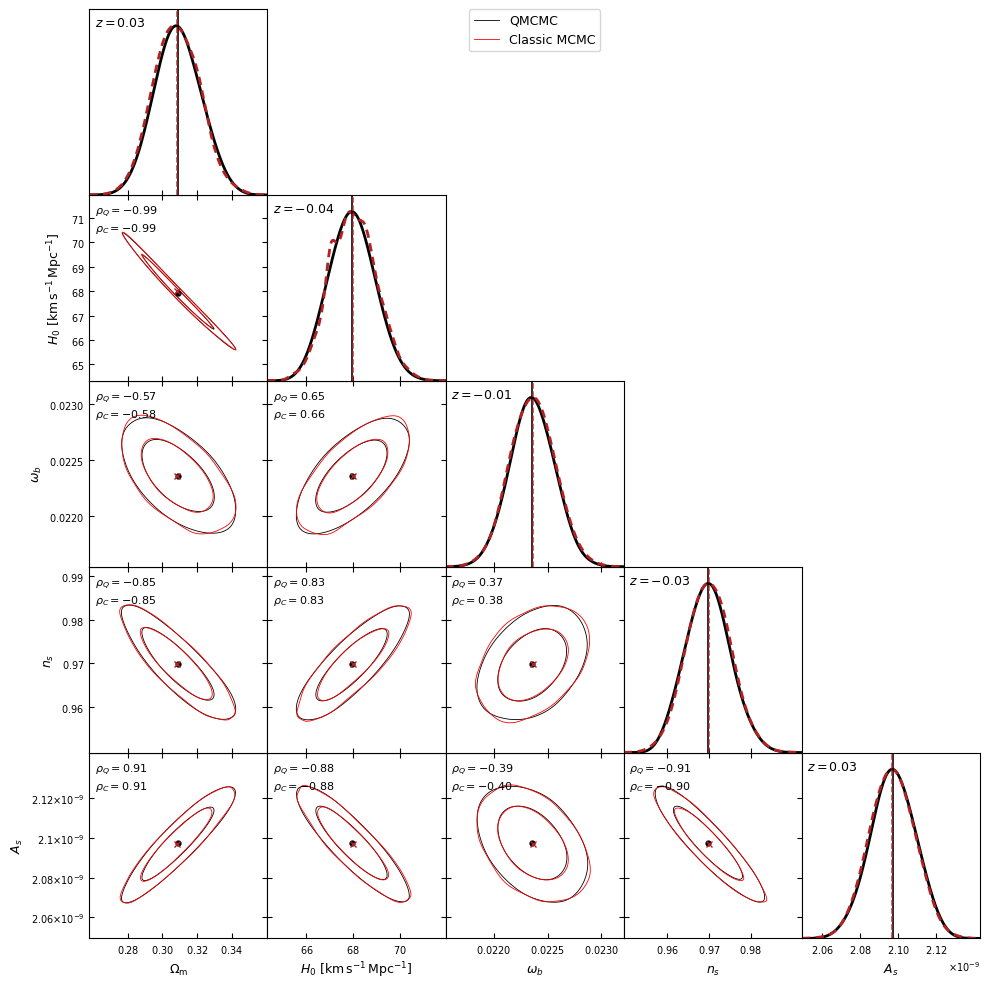}
\includegraphics[width=0.45\hsize]{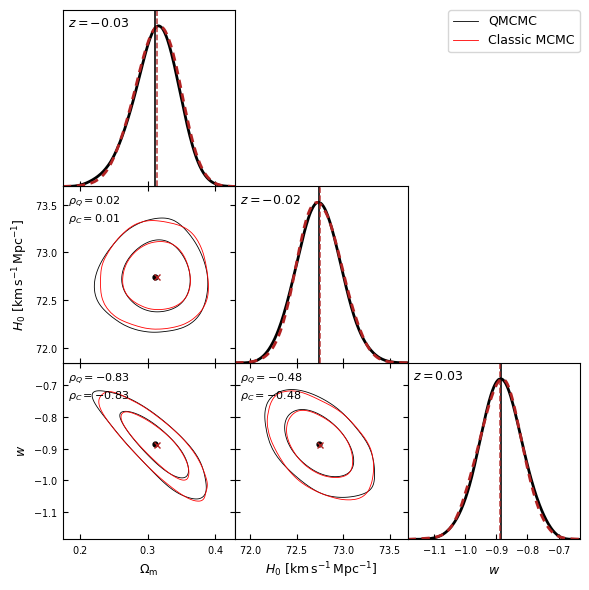}
\caption{$68\%$ and $95\%$ Contour plots from our QMCMC algorithm and the classical MCMC \texttt{emcee} tool. Left panel: for the CMB, considering 5 cosmological parameters for the $\Lambda$CDM. Right panel: for the SNe Ia, considering 3 parameters for the $w$CDM model. Here, also the $z$-score and $\rho$ for each panel are shown.}
\label{Fig_QMCMC_results}
\end{figure*}

We now show the results obtained for real cosmological functions. We recall that we investigate two simple cosmological models in 3 and 5 dimensions, the former considering a $w$CDM model computing the contours for $w$, $\Omega_M$ and $H_0$, and the latter for the standard  $\Lambda$CDM model where we instead compute the contours for $\Omega_M$, $H_0$, $\omega_B$, $n_s$, and $A_s$. The results are displayed in Fig. \ref{Fig_QMCMC_results} both for our QMCMC and \texttt{emcee}. The number of chains for QMCMC is $15$ in both cases.

For the CMB results shown in the left panel, no explicit prior has been defined. Again, for the comparison between quantum and classical results, the mean points and the contours of the two algorithms overlap almost completely, the $z$-scores are all almost 0, and the correlations between parameters are almost identical in the two runs. This confirms the reliability of the results given by our QMCMC. A similar conclusion can also be derived for the results obtained from the SNe Ia, shown in the right panel. In this case, we have defined the same Gaussian prior for both the quantum and classical routines. In particular:

\begin{itemize}
    \item $\Omega_M$: mean=0.30,  standard deviation=0.05.
    \item $H_0$: mean=73,  standard deviation=2.
    \item $w$:  mean=-1.0,  standard deviation=0.1.
\end{itemize}

The aim is to see if the behaviour of the two algorithms is consistent when we consider the contribution of an external Gaussian prior. Again, the contours almost overlap, meaning that the QMCMC is robust when defining a prior explicitly. 

We now show some information on the performance of the QMCMC runs. For the CMB, the convergence has been reached after 97000 steps (24000 if only $R-1$ was considered), with a total elapsed time of around 19 hours. The final acceptance rate is $20\%$, and the ESS is $800$. This shows a strong autocorrelation inside the chains, confirmed by a mean $\tau=1933$. For the SNe IA case, instead, the convergence has been reached after 6500 steps (4000 if only $R-1$ was considered), with a final acceptance rate of $55 \%$. The final mean ESS is $889$. The final elapsed time is $2092$ seconds.

Regarding the speed of the QMCMC with respect to classical counterparts, here the main bottleneck is the evaluation of the likelihoods functions, which remains classical, and in the transpilation of the quantum circuit on the aersimulator, the latter being the operation of rewriting the circuit to optimize its depth and connectivity in view of quantum hardware (or in this case, the emulator) used. Nevertheless, what is important is the number of steps the chains take to converge to the distributions. This depends on different factors, like the initial step size of a single step, as well as the convergence criteria. For the latter, we are currently using both $\tau$ and $R-1$ for our QMCMC, but usually one can suffice. Indeed, it is worth noting that \texttt{emcee} uses a different philosophy in how the chains evolve, and only $\tau$ as a convergence check. Even so, the results are remarkably consistent. In this sense, one could look for an advantage with respect to classical routines in future analyses.

\section{Conclusions}
In this work, we present the first implementations of a QMCMC algorithm that we built and applied to cosmological problems. Our algorithm proposes the steps of the chains moving in the parameter space via a quantum circuit, then computes the acceptance rate classically. Ideally, the possible advantages of this method with respect to a classical algorithm are mainly in the fact that the Quantum Circuit proposes steps of variable length, considering both the real and imaginary parts of the resulting statevector, an object which is not found in classical computations. This could provide a degree of variability in the exploration of the parameter space that could help the convergence of the algorithm. 

We have used our algorithm first with a test function, and then with two cosmological cases, considering the $\chi^2$ functions associated with SNe Ia and CMB. Being the first implementations and analyses performed with this algorithm, the results are encouraging. Indeed, we managed to correctly find the region around the global minimum both for the test function and for the cosmological computations, finding for the latter contours which are almost identical to what can be derived with a reliable classical tool, \texttt{emcee}. 

Possible improvements may be as follows:
\begin{itemize}
\item Automatize the choice of the initial step size for the specific problem to find the optimal configuration for faster convergence.
\item Finding a way to parallelize the computation of the chains for our QMCMC.
\item Testing our algorithm with more complex cosmological computations, increasing even more the dimensions of the parameter space by combining more probes (like combining CMB and SNe Ia, or considering other probes). Integrating our QMCMC with proper cosmological theory codes like CLASS or CAMB is also related to this point.
\item Testing the scalability of our QMCMC with real quantum hardware, assessing how the noise and connectivity affect our circuit, especially once we evaluate more complex cosmological computations. 

\end{itemize}

\section*{Acknowledgment}

This work is supported by the  ICSC - Centro Nazionale di Ricerca in HPC, Big Data e Quantum Computing, CN00000013, Spoke 10 "Quantum Computing", CUP C53C22000350006.

\bibliographystyle{IEEEtran}
\bibliography{IEEEabrv,conference}

\end{document}